\documentclass[letterpaper,twocolumn,10pt]{article}

\usepackage[compact]{titlesec}
\usepackage{enumitem}
\setlist{noitemsep}

\usepackage{usenix}
\usepackage{cite}

\usepackage{hyperref}
\usepackage{amsmath}
\usepackage{stfloats}
\usepackage{url}

\usepackage{algorithm}
\usepackage[noend]{algpseudocode}
\usepackage[table]{xcolor}
\usepackage{dirtytalk}
\usepackage{amsfonts}
\usepackage{amsthm}
\usepackage{makecell}
\usepackage{listings}
\usepackage{xspace}
\usepackage{graphicx}
\usepackage{tikz}
\usepackage{pgfplots}
\usepackage{pgfplotstable}
\pgfplotsset{compat=1.14}
\usepackage{booktabs}
\usepackage{ctable}
\usepackage[warn]{textcomp}

\usepackage{microtype}

\usepackage{etoolbox}
\makeatletter
\patchcmd{\@makecaption}
  {\scshape}
  {}
  {}
  {}
\makeatother

\definecolor{dkgreen}{rgb}{0,0.6,0}
\definecolor{gray}{rgb}{0.5,0.5,0.5}
\definecolor{mauve}{rgb}{0.58,0,0.82}

\newcommand{\progn}[1]{\texttt{#1}\xspace}
\newcommand{\system}[1]{#1\xspace}
\newcommand{\funcn}[1]{\texttt{#1}\xspace}
\newcommand{\inlinecode}[1]{\texttt{#1}\xspace}

\newcommand{\ID}[1]{\textit{#1}}

\newcommand{\SHRIKE}{\textsc{Shrike}\xspace}
\newcommand{\SIEVE}{\textsc{Sieve}\xspace}
\newcommand{\PHP}{\system{PHP}}
\newcommand{\Python}{\system{Python}}
\newcommand{\Ruby}{\system{Ruby}}
\newcommand{\dlmalloc}{\progn{dlmalloc}}
\newcommand{\tcmalloc}{\progn{tcmalloc}}

\newcommand{\avrlibc}{\progn{avrlibc}}

\newcommand{\Windows}{\system{Windows}}

\newcommand{\malloc}{\funcn{malloc}}
\newcommand{\realloc}{\funcn{realloc}}
\newcommand{\calloc}{\funcn{calloc}}
\newcommand{\free}{\funcn{free}}

\algnewcommand{\LineComment}[1]{\State \(\triangleright\) #1}

\theoremstyle{definition}

\graphicspath{ {images/} }

\begin{document}
\date{}

\title{\Large \bf Automatic Heap Layout Manipulation for
Exploitation}


\author{
{\rm Sean Heelan}\\
{\rm sean.heelan@cs.ox.ac.uk}\\
University of Oxford
\and
{\rm Tom Melham}\\
{\rm tom.melham@cs.ox.ac.uk}\\
University of Oxford
\and
{\rm Daniel Kroening}\\
{\rm kroening@cs.ox.ac.uk}\\
University of Oxford
} 

\maketitle

\thispagestyle{empty}

\subsection*{Abstract}

Heap layout manipulation is integral to exploiting heap-based memory
corruption vulnerabilities.  In this paper we present the first automatic
approach to the problem, based on pseudo-random black-box search.  Our
approach searches for the inputs required to place the source of a
heap-based buffer overflow or underflow next to heap-allocated objects that
an exploit developer, or automatic exploit generation system, wishes to read
or corrupt.  We present a framework for benchmarking heap layout
manipulation algorithms, and use it to evaluate our approach on several
real-world allocators, showing that pseudo-random black box search can be
highly effective.  We then present \SHRIKE, a novel system that can perform
automatic heap layout manipulation on the PHP interpreter and can be used in
the construction of control-flow hijacking exploits.  Starting from PHP's
regression tests, \SHRIKE discovers fragments of PHP code that interact with
the interpreter's heap in useful ways, such as making allocations and
deallocations of particular sizes, or allocating objects containing
sensitive data, such as pointers.  \SHRIKE then uses our search algorithm to
piece together these fragments into programs, searching for one that
achieves a desired heap layout.  \SHRIKE allows an exploit developer to focus
on the higher level concepts in an exploit, and to defer the resolution of
heap layout constraints to \SHRIKE.  We demonstrate this by using \SHRIKE in
the construction of a control-flow hijacking exploit for the PHP
interpreter.

\section{Introduction}
\label{sec:introduction}
Over the past decade several researchers~\cite{Avgerinos11AEG, Brumley08APEG, Heelan09Msc, mayhem12} have addressed the problem of automatic exploit generation (AEG) for stack-based buffer overflows. These papers describe algorithms for automatically producing a control-flow hijacking exploit, under the assumption that an input is provided, or discovered, that results in the corruption of an instruction pointer stored on the stack. However, stack-based buffer overflows are just one type of vulnerability found in software written in C and C++. Out-of-bounds (OOB) memory access from heap buffers is a common flaw and, up to now, has received little attention in terms of automation. Heap-based memory corruption differs significantly from stack-based memory corruption. In the latter case the data that the attacker may corrupt is limited to whatever is on the stack and can be varied by changing the execution path used to trigger the vulnerability. For heap-based corruption, it is the physical layout of dynamically allocated buffers in memory that determines what gets corrupt:ed. The attacker must reason about the heap layout to automatically construct an exploit.  In~\cite{repel17}, exploits for heap-based vulnerabilities are considered, but the foundational problem of producing inputs that guarantee a particular heap layout is not addressed.

To leverage OOB memory access as part of an exploit, an attacker will usually want to position some dynamically allocated buffer 
$D$, the OOB access destination, relative to some other dynamically allocated
buffer $S$, the OOB access source.\footnote{Henceforth, when we refer to the `source' and `destination' we mean the source or destination buffer of the overflow or underflow.} The desired positioning will depend on whether
the flaw to be leveraged is an overflow or an underflow, and on the
control the attacker has over the offset from $S$ that will be accessed. Normally, the attacker wants to position $S$ and $D$ so that, when the vulnerability is triggered, $D$ is corrupted while minimising collateral damage to other heap allocated structures.

Allocators do not expose an API to allow a user to
control relative positioning of allocated memory regions. In fact, the ANSI C specification~\cite{ANSI:c89} explicitly states 

\begin{quotation}
\noindent\textit{The order and contiguity of storage allocated by successive calls to the calloc, malloc, and realloc functions is unspecified.}
\end{quotation}

\noindent Furthermore, applications that use dynamic memory allocation do not expose an API allowing an attacker to directly interact with the allocator in an arbitrary manner. An exploit developer must first discover the allocator interactions that can be indirectly triggered via the application's API, and then leverage these to solve the layout problem. In practice, both problems are usually solved manually; this requires expert knowledge of the internals of both the heap allocator and the application's use of it. 

\subsection{An Example}
\begin{lstlisting}[float, language=C, caption={Example API offered by a target program. }, label={lst:example_code}]
typedef struct {
    DisplayFn display;
    char *n;
    unsigned *id
} User;

User* create(char *name) {
    if (!strlen(name) || strlen(name) >= 8) 
        return 0;
    User *user = malloc(sizeof(User));
    user->display = &printf;
    user->n = malloc(strlen(name) + 1);
    strlcpy(user->n, name, 8);
    user->id = malloc(sizeof(unsigned));
    get_uuid(user->id);
    return user;
}

void destroy(User *user) {
    free(user->id);
    free(user->n);
    free(user);
}

void rename(User *user, char *new) {
    strlcpy(user->n, new, 12);
}

void display(User *user) {
    user->display(user->n);
}
\end{lstlisting}

Consider the code in Listing~\ref{lst:example_code} showing the API for a target program. The \inlinecode{rename} function contains a heap-based overflow if the new name is longer than the old name. One way for an attacker to exploit the flaw in the \inlinecode{rename} function is to try to position a buffer allocated to hold the \inlinecode{name} for a \inlinecode{User} immediately before a \inlinecode{User} structure. The \inlinecode{User} structure contains a function pointer as its first field and an attacker in control of this field can redirect the control flow of the target to a destination of their choice by then calling the \inlinecode{display} function.

As the attacker cannot directly interact with the allocator, the desired heap layout must be achieved indirectly utilising those functions in the target's API which perform allocations and deallocations. While the \inlinecode{create} and \inlinecode{destroy} functions do allow the attacker to make allocations and deallocations of a controllable size, other allocator interactions that are unavoidable also take place, namely the allocation and deallocation of the buffers for the \inlinecode{User} and \inlinecode{id}. We refer to these unwanted interactions as \emph{noise}, and such interactions, especially allocations, can increase the difficulty of the problem by placing buffers between the source and destination. 

\begin{figure}
\centering
\includegraphics[scale=.46]{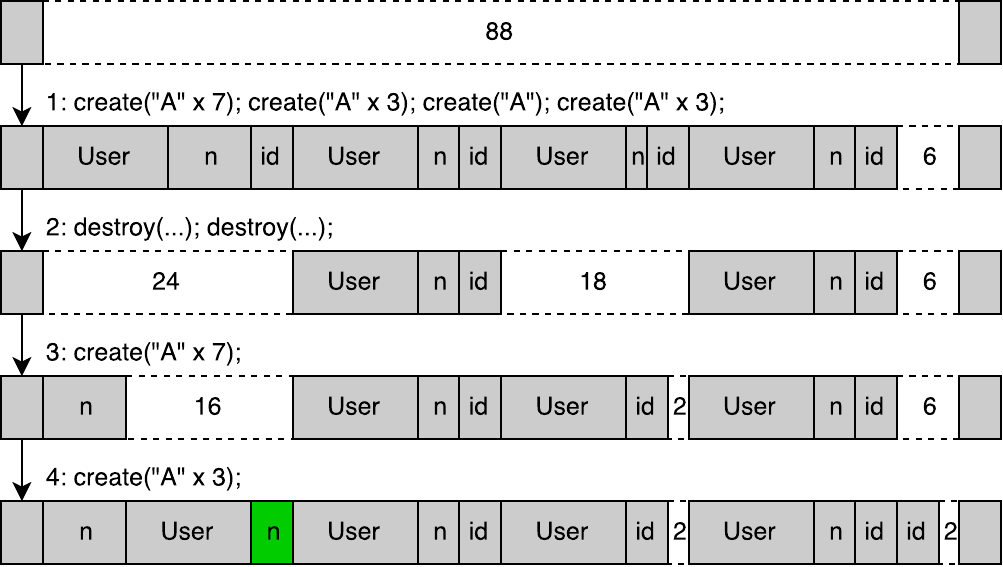}
\caption{An series of interactions which result in a name buffer immediately prior to a \inlinecode{User} structure.}
\label{fig:victim_and_allocator}
\end{figure}

Figure~\ref{fig:victim_and_allocator} shows one possible sequence in which the \inlinecode{create} and \inlinecode{destroy} functions are used to craft the desired heap layout.\footnote{Assume a best-fit allocator using last-in-first-out free lists to store free chunks, no limit on free chunk size, no size rounding and no inline allocator metadata. Furthermore, assume that pointers are 4 bytes in size and that a \inlinecode{User} structure is 12 bytes in size.} The series of interactions performed by the attacker are as follows:
\begin{enumerate}
  \item Four users are created with names of length 7, 3, 1, and 3 letters, respectively.
  \item The first and the third user are destroyed, creating two holes: One of size 24 and one of size 18.
  \item A user with a name of length 7 is created. The allocator uses the hole of size 18 to satisfy the allocation request for the 12-byte \inlinecode{User} structure, leaving 6 free bytes. The request for the 8-byte name buffer is satisfied using the 24-byte hole, leaving a hole of 16 bytes. An allocation of 4 bytes for the \inlinecode{id} then reduces the 6 byte hole to 2. 
  \item A user with a name of length 3 is created. The 16-byte hole is used for the \inlinecode{User} object, leaving 4 bytes into which the name buffer is then placed. This results in the name buffer, highlighted in green, being directly adjacent to a \inlinecode{User} structure.
\end{enumerate}

Once this layout has been achieved an overflow can be triggered using the \inlinecode{rename} function, corrupting the \inlinecode{display} field of the \inlinecode{User} object. The control flow of the application can then be hijacked by calling the \inlinecode{display} function with the corrupted \inlinecode{User} object as an argument.

\subsection{Contributions}
Our contributions are as follows:

\begin{enumerate}
\item An analysis of the heap layout manipulation (HLM) problem as a standalone task within the context of automatic exploit generation, outlining its essential aspects and describing the factors which influence its complexity.
\item \SIEVE, an open source framework for constructing benchmarks for heap layout manipulation and evaluating algorithms.
\item A pseudo-random black box search algorithm for heap layout manipulation. Using \SIEVE, we evaluate the effectiveness of this algorithm on three real-world allocators, namely \dlmalloc, \avrlibc and \tcmalloc.
\item An architecture, and proof-of-concept implementation, for a system that integrates automatic HLM into the exploit development process. The implementation, \SHRIKE, automatically solves heap layout constraints that arise when constructing exploits for the \PHP interpreter. \SHRIKE also demonstrates a novel approach to integrating an automated reasoning engine into the exploit development process. The exploit developer produces a partial exploit with markers indicating heap layout problems to be solved. \SHRIKE takes this partial exploit as input and completes it by solving these problems.
\end{enumerate}

The source code for \SHRIKE and \SIEVE can be found at \url{https://sean.heelan.io/heaplayout}.

\section{The Heap Layout Manipulation Problem in Deterministic Settings}
\label{sec:the_heap_layout_optimisation_problem}
As of 2018, the most common approach to solving heap layout manipulation problems is manual work by experts. An analyst examines the allocator's implementation to gain an understanding of its internals; then, at run-time, they inspect the state of its various data structures to determine what interactions are necessary in order to manipulate the heap into the required layout. 

Heap layout manipulation primarily consists of two activities: creating and filling \emph{holes} in memory. A hole is a free area of memory that the allocator may use to service future allocation requests. Holes are filled to force the positioning of an allocation of a particular size elsewhere, or the creation of a fresh area of memory under the management of the allocator. Holes are created to capture allocations that would otherwise interfere with the layout one is trying to achieve. This process is documented in the literature of the hacking and computer security communities, with a variety of papers on the internals of individual allocators~\cite{Phrack:MaXX, Phrack:AnonFree, Phrack:jp, Phrack:ArgpJeMalloc}, as well as the manipulation and exploitation of those allocators when embedded in applications~\cite{SolarDes:Unlink, Phrack:ArgpVLC, Phrack:ArgpFF}. 

The process is complicated by the fact that -- when constructing an exploit -- one cannot directly interact with the allocator, but instead must use the API exposed by the target program. Manipulating the heap state via the program's API is often referred to as \emph{heap feng shui} in the computer security literature~\cite{Sotirov07HeapFengShui}. Discovering the relationship between program-level API calls and allocator interactions is a prerequisite for real-world HLM but can be addressed separately, as we demonstrate in section~\ref{sec:php_experiments}.

\subsection{Problem Restrictions for a Deterministic Setting}

\begin{figure}
\centering
\includegraphics[scale=.5]{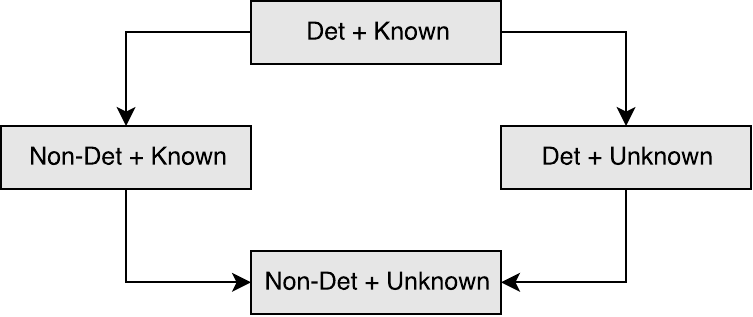}
\caption{The challenges in achieving a particular layout vary depending on whether the allocator behaves deterministically or non-deterministically and whether or not the starting state of the heap is known.}
\label{fig:problem_variants}
\end{figure}

There are four variants of the HLM problem, as shown in Figure~\ref{fig:problem_variants}, depending on whether the allocator is deterministic or non-deterministic and whether the starting state is known or unknown. A deterministic allocator is one that does not utilise any random behaviour when servicing allocation requests. The majority of allocators are deterministic, but some, such as the \Windows system allocator, \texttt{jemalloc} and the \textsc{DieHard} family of allocators~\cite{DieHard, DieHarder}, do utilise non-determinism to make exploitation more difficult. The starting state of the heap at which the attacker can begin interacting with the allocator is given the allocations and frees that have taken place up to that point. For the starting state to be known, this sequence of interactions must be known. 

In this paper we consider a known starting state and a deterministic allocator, and assume there are no other actors interacting with the heap. While restricted, this both corresponds to a set of real world exploitation scenarios and provides a building block for addressing the other three problem variants.

Local privilege escalation exploits are a scenario in which these restrictions are usually met, as the attacker can often tell what allocations and deallocations take place prior to their interactions. For remote and client-side targets, the starting state is usually not known. However, for some such targets it is possible to force the creation of a new heap in a predictable state.

When unknown starting states and non-determinism must be dealt with, approaches such as allocating a large number of objects on the heap in the hope of corrupting one when the vulnerability is triggered are often used. However, in the problem variant we address it is usually possible to position the overflow source relative to a \emph{specific} target buffer. Thus our objective in this variant of the HLM problem is as follows:

\begin{quotation}
\noindent\textit{Given the API for a target program and a means by which to allocate a source and destination buffer, find a sequence of API calls that position the destination and source at a specific offset from each other.}
\end{quotation}

\subsection{Challenges}
\label{sec:challenges}

There are several challenges that arise when trying to perform HLM and when trying to construct a general, automated solution. In this section we outline those that are most likely to be significant.  


\subsubsection{Interaction Noise}

Before continuing we first must informally define the concept of an `\emph{interaction sequence}': an allocator \emph{interaction} is a call to one of its allocation or deallocation functions, while an \emph{interaction sequence} is a list of one or more interactions that result from the invocation of a function in the target program's API. As an attacker cannot directly invoke functions in the allocator they must manipulate the heap via the available interaction sequences. As an example, when the \inlinecode{create} function from Listing~\ref{lst:example_code} is called the resulting interaction sequence consists of three interactions in the form of the three calls to \inlinecode{malloc}. The \inlinecode{destroy} function also provides an interaction sequence of length three, in this case consisting of three calls to \inlinecode{free}.

For a given interaction sequence there can be interactions that are beneficial, and assist with manipulation of the heap into a layout that is desirable, and also interactions that are either not beneficial (but benign), or in fact are detrimental to the heap state in terms of the layout one is attempting to achieve. We deem those interactions that are not actively manipulating the heap into a desirable state to be \emph{noise}.

For example, the \inlinecode{create} function from Listing~\ref{lst:example_code} provides the ability to allocate buffers between 2 and 8 bytes in size by varying the length of the \inlinecode{name} parameter. However, two other unavoidable allocations also take place -- one for the \inlinecode{User} structure and one for the \inlinecode{id}. As shown in Figure~\ref{fig:victim_and_allocator}, some effort must be invested in crafting the heap layout to ensure that the noisy \inlinecode{id} allocation is placed out of the way and a \inlinecode{name} and \inlinecode{User} structure end up next to each other. 


\subsubsection{Constraints on Allocator Interactions}

An attacker's access to the allocator is limited by what is allowed by the program they are interacting with. The interface available may limit the sizes that may be allocated, the order in which they may be allocated and deallocated, and the number of times a particular size may be allocated or deallocated. Depending on the heap layout that is desired, these constraints may make the desired layout more complex to achieve, or even impossible.





\subsubsection{Diversity of Allocator Implementations}

The open ended nature of allocator design and implementation means any approach that involves the production of a formal model of a particular allocator is going to be costly and likely limited to a single allocator, and perhaps even a specific version of that allocator. While \avrlibc is a mere 350 lines of code, most of the other allocators we consider contain thousands or tens of thousands of lines of code. Their implementations involve complex data structures, loops without fixed bounds, interaction with the operating system and other features that are often terminally challenging for semantics-aware analyses, such as model checking and symbolic execution. A detailed survey of the data structures and algorithms used in allocators is available in~\cite{Wilson:DynamicStorageSurvey95}.

\subsubsection{Interaction Sequence Discovery}

Since in most situations one cannot directly interact with the allocator, an attacker needs to discover what interaction sequences with the allocator can be indirectly triggered via the program's API. This problem can be addressed separately to the main HLM problem, but it is a necessary first step. In section~\ref{sec:php_experiments} we discuss how we solved this problem for the \PHP language interpreter.

\section{Automatic Heap Layout Manipulation}
\label{sec:solution}
We now present our pseudo-random black box search algorithm for HLM, and two evaluation frameworks we have embedded it in to solve heap layout problems on both synthetic benchmarks and real vulnerabilities. The algorithm is theoretically and practically straightforward. There are two strong motivations for initially avoiding complexity. 

Firstly, there is no existing prior work on automatic HLM and a straightforward algorithm provides a baseline that future, more sophisticated, implementations can be compared against if necessary.

Secondly, despite the potential size of the problem measured by the number of possible combinations of available interactions, there is significant symmetry in the solution space for many problem instances. Since our measure of success is based on the relative positioning of two buffers, large equivalence classes of solutions exist as:
\begin{enumerate}
	\item Neither the absolute location of the two buffers, nor their relative position to other buffers, matters.
   \item The order in which holes are created or filled usually does not matter.
\end{enumerate}


It is often possible to solve a layout problem using significantly differing input sequences. Due to these solution space symmetries, we propose that a pseudo-random black box search could be a solution for a sufficiently large number of problem instances as to be worthwhile.

To test this hypothesis, and demonstrate its feasibility on real targets, we constructed two systems. The first, described in section~\ref{sec:hlo_standalone} allows for synthetic benchmarks to be constructed with any allocator exposing the standard ANSI interface for dynamic memory allocation. The second system, described in section~\ref{sec:hlo_php}, is a fully automated HLM system designed to work with the \PHP interpreter. 

\subsection{\SIEVE: An Evaluation Framework for HLM Algorithms}
\label{sec:hlo_standalone}

To allow for the evaluation of search algorithms for HLM across a diverse array of benchmarks we constructed \SIEVE. It allows for flexible and scalable evaluation of new search algorithms, or testing existing algorithms on new allocators, new interaction sequences or new heap starting states. There are two components to \SIEVE:

\begin{enumerate}
    \item The \SIEVE driver which is a program that can be linked with any allocator exposing the \malloc, \free, \calloc and \realloc functions. As input it takes a file specifying a series of allocation and deallocation requests to make, and produces as output the distance between two particular allocations of interest. Allocations and deallocations are specified via directives of the following forms: 
    
    \begin{enumerate}
    	\item \inlinecode{<malloc size ID>}
    	\item \inlinecode{<calloc nmemb size ID>}
    	\item \inlinecode{<free ID>}
    	\item \inlinecode{<realloc oldID size ID>}
    	\item \inlinecode{<fst size>}
    	\item \inlinecode{<snd size>}
    \end{enumerate}

	Each of the first four directives are translated into an invocation of their corresponding memory management function, with the \inlinecode{ID} parameters providing an identifier which can be used to refer to the returned pointers from \malloc, \calloc and \realloc, when they are passed to \free or \realloc. The final two directives indicate the allocation of the two buffers that we are attempting to place relative to each other. We refer to the addresses that result from the corresponding allocations as $addrFst$ and $addrSnd$, respectively. After the allocation directives for these buffers have been processed, the value of $(addrFst - addrSnd)$ is produced.
	
    \item The \SIEVE framework which provides a Python API for running HLM experiments. It has a variety of features for constructing candidate solutions, feeding them to the driver and retrieving the resulting distance, which are explained below. This functionality allows one to focus on creating search algorithms for HLM. 
\end{enumerate}

\begin{algorithm}[t]
\caption{Find a solution that places two allocations in memory at a
specified distance from each other.  The integer $g$ is the number of
candidates to try, $d$ the required distance, $m$~the maximum candidate size
and $r$ the ratio of allocations to deallocations for each candidate.}
\label{alg:search}
\begin{algorithmic}[1]
\Function{Search}{$g, d, m, r$}
\For{$i \gets 0, g - 1$}
    \State $\ID{cand} \gets \ID{ConstructCandidate}(m, r)$
    \State $\ID{dist} \gets \ID{Execute}(\ID{cand})$
    \If{$\ID{dist} = d$}
        \State \Return $\ID{cand}$
    \EndIf
\EndFor
\State \Return $\ID{None}$
\EndFunction
\medskip

\Function{ConstructCandidate}{$m, r$}
\State $\ID{cand} \gets \ID{InitCandidate}(\ID{GetStartingState}())$
\State $\ID{len} \gets \ID{Random}(1, m)$
\State $\ID{fstIdx} \gets \ID{Random}(0, \ID{len} - 1)$
\For{$i \gets 0, \ID{len} - 1$}
\If{$i = \ID{fstIdx}$}
    \State $\ID{AppendFstSequence}(\ID{cand})$
\ElsIf{$\ID{Random}(1, 100) \leq r$}
    \State $\ID{AppendAllocSequence}(\ID{cand})$
\Else
    \State $\ID{AppendFreeSequence}(\ID{cand})$
\EndIf
\EndFor
\State $\ID{AppendSndSequence}(\ID{cand})$
\State \Return $\ID{cand}$
\EndFunction
\end{algorithmic}
\end{algorithm}

We implemented a pseudo-random search algorithm for HLM on top of \SIEVE, and it is shown as Algorithm~\ref{alg:search}. The $m$ and $r$ parameters are what make the search \emph{pseudo-random}. While one could potentially use a completely random search, it makes sense to guide it away from candidates that are highly unlikely to be useful due to extreme values for $m$ and $r$. There are a few points to note on the \SIEVE framework's API in order to understand the algorithm:

\begin{itemize}
    \item The directives to be passed to the driver are represented in the framework via a \inlinecode{Candidate} class. The \inlinecode{InitCandidate} function creates a new \inlinecode{Candidate}. 
    
    \item Often one may want to experiment with performing HLM after a number of allocator interactions, representing initialisation of the target application \emph{before} the attacker can interact, have taken place. \SIEVE can be configured with a set of such interactions that can be retrieved via the \inlinecode{GetStartingState} function. \inlinecode{InitCandidate} can be provided with the result of \inlinecode{GetStartingState} (line 9).
    
    \item The available interaction sequences impact the difficulty of HLM, i.e. if an attacker can trigger individual allocations of arbitrary sizes they will have more precise control of the heap layout than if they can only make allocations of a single size. To experiment with changes in the available interaction sequences, the user of \SIEVE overrides the \inlinecode{AppendAllocSequence} and \inlinecode{AppendFreeSequence}\footnote{\inlinecode{AppendFreeSequence} function will detect if there are no allocated buffers to free and redirect to \inlinecode{AppendAllocSequence} instead.} functions to select one of the available interaction sequences and append it to the candidate (lines 16-18).
    
    \item The directive to allocate the first buffer of interest is placed at a random offset within the candidate (line 14), with the directive to allocate the second buffer of interest placed at the end (line 19). To experiment with the addition of noise in the allocation of these buffers, the \inlinecode{AppendFstSequence} and \inlinecode{AppendSndSequence} functions can be overloaded.
    
    \item The \inlinecode{Execute} function takes a candidate, serialises it into the form required by the \SIEVE driver, executes the driver on the resulting file and returns the distance output by the driver (line 4).
    
    \item As the value output by the driver is $(addrFst - addrSnd)$, to search for a solution placing the buffer allocated first \emph{before} the buffer allocated second, a negative value can be provided for the $d$ parameter to \inlinecode{Search}. Providing a positive value will search for a solution placing the buffers in the opposite order. In this manner overflows and underflows can be simulated, with either temporal order of allocation for the source and destination (line 5). 
\end{itemize}

The experimental setup used to evaluate pseudo-random search as a means for solving HLM problems on synthetic benchmarks is described in section~\ref{sec:synth_experiments}.

\subsection{\SHRIKE: A HLM System for PHP}
\label{sec:hlo_php}

For real-world usage the search algorithm must be embedded in a system that solves a variety of other problems in order to allow the search to take place. To evaluate the feasibility of end-to-end automation of HLM we constructed \SHRIKE, a HLM system for the \PHP interpreter. We choose \PHP as it has a number of attributes that make it ideal for experimentation. \PHP combines a large, modern application containing complex functionality, with a language that is relatively stable and easy to work with in an automated fashion. On top of that, it has an open version control system and bug tracker.

Furthermore, \PHP is an interesting target from a security point of view as the ability to exploit heap-based vulnerabilities locally in \PHP allows attackers to increase their capabilities in situations where the \PHP environment has been hardened~\cite{Esser09PHP}.

\begin{figure}
\centering
\includegraphics[scale=.55]{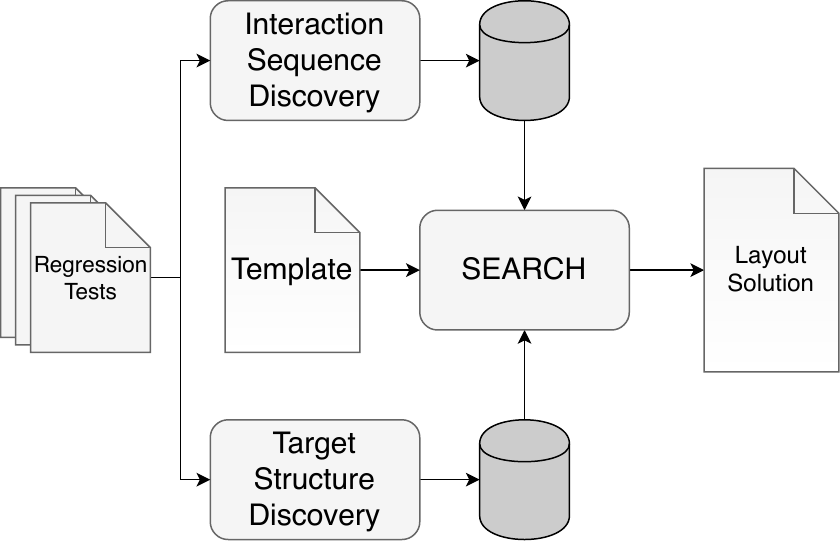}
\caption{Architecture diagram for \SHRIKE}
\label{fig:shrike_arch}
\end{figure}

The architecture of \SHRIKE is shown in Figure~\ref{fig:shrike_arch}. We implemented the system as three distinct phases: 

\begin{itemize}
    \item A component that identifies fragments of \PHP code that provide distinct allocator interaction sequences (Section~\ref{sec:identifying_available_interaction_sequences}).
    \item A component that identifies dynamically allocated structures that may be useful to corrupt or read as part of an exploit, and a means to trigger their allocation (Section~\ref{sec:automatic_identification_of_target_structures}).
    \item A search procedure that pieces together the fragments triggering allocator interactions to produce \PHP programs as candidates (Section~\ref{sec:search}). The user specifies how to allocate the source and destination, as well as how to trigger the vulnerability, via a template (Section~\ref{sec:specifying_candidate_structure}). 
\end{itemize}

The first two components can be run once and the results stored for use during the search. If successful, the output of the search is a new \PHP program that manipulates the heap to ensure that when the specified vulnerability is triggered the source and destination buffers are adjacent. 

To support the functionality required by \SHRIKE we implemented an extension for \PHP. This extension provides functions that can be invoked from a \PHP script to enable a variety of features including recording the allocations that result from invoking a fragment of \PHP code, monitoring allocations for the presence of \emph{interesting} data, and checking the distance between two allocations. We carefully implemented the functionality of this extension to ensure that it does not modify the heap layout of the target program in any way that would invalidate search results. However, all results are validated by executing the solutions in an unmodified version of \PHP.

\subsubsection{Identifying Available Interaction Sequences}
\label{sec:identifying_available_interaction_sequences}

To discover the available interaction sequences it is necessary to construct self-contained fragments of \PHP code and determine the allocator interactions each fragment triggers. Correlating code fragments with the resulting allocator interactions is straightforward: we instrument the \PHP interpreter to record the allocator interactions that result from executing a given fragment. Constructing valid fragments of \PHP code that trigger a diverse set of allocator interactions is more involved. 

We resolve the latter problem by implementing a fuzzer for the \PHP interpreter that leverages the regression tests that come with \PHP, in the form of \PHP programs. This idea is based on previous work that used a similar approach for the purposes of vulnerability detection~\cite{holler12Langfuzz, heelan14ghosts}. The tests provide examples of the functions that can be called, as well as the number and types of their arguments. The fuzzer then mutates existing fragments, to produce new fragments with new behaviours.

To tune the fuzzer towards the discovery of fragments that are useful for HLM, as opposed to vulnerability discovery, we made the following modifications:

\begin{itemize}
    \item We use mutations that are intended to produce an interaction sequence that we have not seen before, rather than a crash. For example, fuzzers will often replace integers with values that may lead to edge cases, such as 0, $2^{32}-1$, $2^{31}-1$ and so on. We are interested in triggering unique allocator interactions however, and so we predominantly mutate tests using integers and string lengths that relate to allocation sizes we have not previously seen.    
    \item Our measure of \emph{fitness} for a generated test is not based on code coverage, as is often the case with vulnerability detection, but is instead based on whether a new allocator interaction sequence is produced, and the length of that interaction sequence. 
    \item We discard any fragments that result in the interpreter exiting with an error.
    \item We favour the shortest, least complex fragments with priority being given to fragments consisting of a single function call. 
\end{itemize} 

\begin{lstlisting}[float, language=PHP, caption={Source for a \PHP test program.}, label={lst:php_test}]
$image = imagecreatetruecolor(180, 30);
imagestring($image, 5, 10, 8, "Text", 0x00ff00);
$gaussian = array(
    array(1.0, 2.0, 1.0),
    array(2.0, 4.0, 2.0)
);
var_dump(imageconvolution($image, $gaussian, 16, 0));
\end{lstlisting}

\begin{lstlisting}[float, language=PHP, caption={The function fuzzing specifications produced from parsing Listing~\ref{lst:php_test}. The letters replacing the function arguments indicate their types. `R' for a resource, `I' for an integer, `F' for a float and `T' for text.}, label={lst:fuzzing_templates}]
imagecreatetruecolor(I, I)
imagestring(R, I, I, I, T, I)
array(F, F, F)
array(R, R)
var_dump(R)
imageconvolution(R, R, I, I)
\end{lstlisting}

As an example, lets discuss how the regression test in Listing~\ref{lst:php_test} would be used to discover interaction sequences. From the regression test the fuzzing specification shown in Listing~\ref{lst:fuzzing_templates} is automatically produced. Fuzzing specifications indicate the name of functions that can be called, along with the types of their arguments. \SHRIKE then begins to fuzz the discovered functions, using the specifications to ensure the correct types are provided for each argument. For example, the code fragments \inlinecode{\$x = imagecreatetruecolor(1, 1)}, \inlinecode{\$x = imagecreatetruecolor(1, 2)}, \inlinecode{\$x = imagecreatetruecolor(1, 3)} etc. might be created and executed to determine what, if any, allocator interactions they trigger.

The output of this stage is a mapping from fragments of \PHP code to a summary of the allocator interaction sequences that occur as a result of executing that code. The summary includes the number and size of any allocations, and whether the sequence triggers any frees.

\subsubsection{Automatic Identification of Target Structures}
\label{sec:automatic_identification_of_target_structures}

In most programs there is a diverse set of dynamically allocated structures that one could corrupt or read to violate some security property of the program. These targets may be program-specific, such as values that guard a sensitive path; or they may be somewhat generic, such as a function pointer. Identifying these targets, and how to dynamically allocate them, can be a difficult manual task in itself. To further automate the process we implemented a component that, as with the fuzzer, splits the \PHP tests into standalone fragments and then observes the behaviour of these fragments when executed. If the fragment dynamically allocates a buffer and writes what appears to be a pointer to that buffer, we consider the buffer to be an interesting corruption target and store the fragment. The user can indicate in the template which of the discovered corruption targets to use, or the system can automatically select one.

\subsubsection{Specifying Candidate Structure}
\label{sec:specifying_candidate_structure}

Different vulnerabilities require different setup in order to trigger e.g. the initialisation of required objects or the invocation of multiple functions. To avoid hard-coding vulnerability-specific information in the candidate creation process, we allow for the creation of candidate templates that define the structure of a candidate. A template is a normal \PHP program with the addition of directives starting with \inlinecode{\#X-SHRIKE}\footnote{As the directives begin with a `\#' they will be interpreted by the normal \PHP interpreter as a comment and thus can be run in both our modified interpreter and an unmodified one.}. The template is processed by \SHRIKE and the directives inform it how candidates should be produced and what constraints they must satisfy to solve the HLM problem. The supported directives are:

\begin{itemize}
    \item \inlinecode{<HEAP-MANIP [sizes]>} Indicates a location where \SHRIKE can insert heap-manipulating sequences. The \inlinecode{sizes} argument is an optional list of integers indicating the allocation sizes that the search should be restricted to.
    \item \inlinecode{<RECORD-ALLOC offset id>} Indicates that \SHRIKE should inject code to record the address of an allocation and associate it with the provided \inlinecode{id} argument. The \inlinecode{offset} argument indicates the allocation to record. Offset 0 is the very next allocation, offset 1 the one after that, and so on. 
    \item \inlinecode{<REQUIRE-DISTANCE idx idy dist>} Indicates that \SHRIKE should inject code to check the distance between the pointers associated with the provided IDs. Assuming $x$ and $y$ are the pointers associated with $idx$ and $idy$ respectively, then if $(x - y = dist)$ \SHRIKE will report the result to the user, indicating this particular HLM problem has been solved. If $(x - y \neq dist)$ then the candidate will be discarded and the search will continue. 
\end{itemize}

\begin{lstlisting}[float, language=PHP, caption={Exploit template for CVE-2013-2110}, label={lst:exploit_template}]
<?php
$quote_str = str_repeat("\xf4", 123);
#X-SHRIKE HEAP-MANIP
#X-SHRIKE RECORD-ALLOC 0 1
$image = imagecreate(1, 2);
#X-SHRIKE HEAP-MANIP 
#X-SHRIKE RECORD-ALLOC 0 2
quoted_printable_encode($quote_str);
#X-SHRIKE REQUIRE-DISTANCE 1 2 0
?>
\end{lstlisting}

A sample template for CVE-2013-2110, a heap-based buffer overflow in \PHP, is shown in Listing~\ref{lst:exploit_template}. In section~\ref{sec:generating_exploit} we explain how this template was used in the construction of a control-flow hijacking exploit for \PHP.  

\subsubsection{Search}
\label{sec:search}

\begin{algorithm}[t]
\caption{Solve the HLM problem described in the provided template~$t$.  The
integer $g$ is the number of candidates to try, $d$ the required distance,
$m$ the maximum number of fragments that can be inserted in place of each
\inlinecode{HEAP-MANIP} directive, and $r$ the ratio of allocations to
deallocation fragments used in place of each \inlinecode{HEAP-MANIP}
directive.}
\label{alg:php_search}
\begin{algorithmic}[1]
\Function{Search}{$t, g, m, r$}
\State $\ID{spec} \gets \ID{ParseTemplate}(t)$
    \For{$i \gets 0, g - 1$}
        \State $\ID{cand} \gets \ID{Instantiate}(\ID{spec}, m, r)$
        \If{$\ID{Execute}(\ID{cand})$}
            \State \textbf{return} $\ID{cand}$
        \EndIf
    \EndFor
    \State \textbf{return} $\ID{None}$
\EndFunction
\medskip

\Function{Instantiate}{$\ID{spec}, m, r$}
    \State $\ID{cand} \gets \ID{NewPHP{\kern0.5pt}Program}()$
    \While{$n \gets \ID{Iterate}(\ID{spec})$}
        \If{$\ID{IsHeapManip}(n)$}
            \State $\ID{code} \gets \ID{GetHeapManipCode}(n, m, r)$
        \ElsIf{$\ID{IsRecordAlloc}(c)$}
            \State $\ID{code} \gets \ID{GetRecordAllocCode}(n)$
        \ElsIf{$\ID{IsRequireDistance}(n)$}
            \State $\ID{code} \gets \ID{GetRequireDistanceCode}(n)$
        \Else
            \State $\ID{code} \gets \ID{GetVerbatim}(n)$
        \EndIf
        \State $\ID{AppendCode}(\ID{cand}, \ID{code})$
    \EndWhile
    \State \Return $\ID{cand}$
\EndFunction
\end{algorithmic}
\end{algorithm}

The search in \SHRIKE is outlined in Algorithm~\ref{alg:php_search}. It takes in a template, parses it and then constructs and executes \PHP programs until a solution is found or the execution budget expires. Candidate creation is shown in the \inlinecode{Instantiate} function. Its first argument is a representation of the template as a series of objects. The objects represent either \SHRIKE directives or normal PHP code and are processed as follows:

\begin{itemize}
    \item The \inlinecode{HEAP-MANIP} directive is handled via the \inlinecode{GetHeapManipCode} function (line 12). The database, constructed as described in section~\ref{sec:identifying_available_interaction_sequences}, is queried for a series of \PHP fragments, where each fragment allocates or frees one of the sizes specified in the \inlinecode{sizes} argument to the directive in the template. If no sizes are provided then all available fragments are considered. If multiple fragments exist for a given size then selection is biased towards fragments with less noise. Between 1 and $m$ fragments are selected and returned. The $r$ parameter controls the ratio of fragments containing allocations to those containing frees. 
    \item The \inlinecode{RECORD-ALLOC} directive is handled via the \inlinecode{GetRecordAllocCode} function (line 14). A \PHP fragment is returned consisting of a call to a function in our extension for \PHP that associates the specified allocation with the specified ID. 
    \item The \inlinecode{REQUIRE-DISTANCE} directive is handled via the \inlinecode{GetRequireDistanceCode} function (line 16). A \PHP fragment is returned with two components. Firstly, a call to a function in our \PHP extension that queries the distance between the pointers associated with the given IDs. Secondly, a conditional statement that prints a success indicator if the returned distance equals the $\ID{distance}$ parameter. 
    \item All code that is not a \SHRIKE directive is included in each candidate verbatim (line 18).
\end{itemize}

The \inlinecode{Execute} function (line 5) converts the candidate into a valid \PHP program and invokes the \PHP interpreter on the result. It checks for the success indicator printed by the code inserted to handle the \inlinecode{REQUIRE-DISTANCE} directive. If that is detected then the solution program is reported. Listing~\ref{lst:php_sample_result} in the appendix shows a solution produced from the template in Listing~\ref{lst:exploit_template}.

\section{Experiments and Evaluation}
\label{sec:experiments}
The research questions we address are as follows:

\begin{itemize}
    \item RQ1: What factors most significantly impact the difficulty of the heap layout manipulation problem in a deterministic setting?
    \item RQ2: Is pseudo-random search an effective approach to heap-layout manipulation?
    \item RQ3: Can heap layout manipulation be automated effectively for real-world programs?
\end{itemize}

We conducted two sets of experiments. Firstly, to investigate the fundamentals of the problem we utilised the system discussed in section~\ref{sec:hlo_standalone} to construct a set of synthetic benchmarks involving differing combinations of heap starting states, interaction sequences, source and destination sizes, and allocators. We chose the \tcmalloc (v2.6.1), \dlmalloc (v2.8.6) and \avrlibc (v2.0) allocators for experimentation. These allocators have significantly different implementations and are used in many real world applications.

An important difference between the allocators used for evaluation is that \tcmalloc (and \PHP) make use of \emph{segregated storage}, while \dlmalloc and \avrlibc do not. In short, for small allocation sizes (e.g. less than a 4KB) segregated storage pre-segments runs of pages into chunks of the same size and will then only place allocations of that size within those pages. Thus, only allocations of the same, or similar, sizes may be adjacent to each other, except for the first and last allocations in the run of pages which may be adjacent to the last or first allocation from other size classes.  

Secondly, to evaluate the viability of our search algorithm on real world applications we ran \SHRIKE on 30 different layout manipulation problems in \PHP. 
All experiments were carried out on a server with 80 Intel Xeon E7-4870 2.40GHz cores and 1TB of RAM, utilising 40 concurrent analysis processes. 

\subsection{Synthetic Benchmarks}
\label{sec:synth_experiments}
\ctable[
    caption = {Synthetic benchmark results after 500,000 candidate solutions generated, averaged across all starting sequences. The full results are in Table~\ref{tab:full_synth} in the appendix. All experiments were run 9 times and the results presented are an average.},
    label = {tab:synth_results},
    notespar,
]{@{}lllllllll@{}}{
}
{
    \FL
    Allocator   & Noise & \makecell{\%\\Overall\\Solved}  & \makecell{\%\\Natural\\Solved} & \makecell{\%\\Reversed\\Solved} 
    \ML
    avrlibc-r2537 & 0 & 100 & 100 & 99 \NN
	dlmalloc-2.8.6 & 0 & 99 & 100 & 98 \NN
	tcmalloc-2.6.1 & 0 & 72 & 75 & 69 \NN
	avrlibc-r2537 & 1 & 51 & 50 & 52 \NN
	dlmalloc-2.8.6 & 1 & 46 & 60 & 31 \NN
	tcmalloc-2.6.1 & 1 & 52 & 58 & 47 \NN
	avrlibc-r2537 & 4 & 41 & 44 & 38 \NN
	dlmalloc-2.8.6 & 4 & 33 & 49 & 17 \NN
	tcmalloc-2.6.1 & 4 & 37 & 51 & 24 \NN
    \LL
}

The goal of evaluation on synthetic benchmarks is to discover the factors influencing the difficulty of problem instances and to highlight the capabilities and limitations of our search algorithm in an environment that we precisely control. The benchmarks were constructed as follows:

\begin{itemize}

\item In real world scenarios it is often the case that the available interaction sequences are noisy. To investigate how varying noise impacts problem difficulty, we constructed benchmarks in which varying amounts of noise are injected during the allocation of the source and destination. In Table~\ref{tab:synth_results}, a value of $N$ in the `Noise' column means that before and after the first allocation of interest, $N$ allocations of size equal to the second allocation of interest allocation are made. 

\item We initialise the heap state prior to executing the interactions from a candidate by prefixing each candidate with a set of interactions. Previous work~\cite{Wilson:DynamicStorageSurvey95} has outlined the drawbacks that arise when using randomly generated heap states to evaluate allocator performance. To avoid these drawbacks we captured the initialisation sequences of \PHP\footnote{\PHP makes use of both the system allocator and its own allocator. We captured the initialisation sequences for both.}, \Python and \Ruby to use in our benchmarks. A summary of the relevant properties of these initialisation sequences can be found in the appendices in table~\ref{tab:init_seq_summary}.

\item As it is not feasible to evaluate layout manipulation for all possible combinations of source and destination sizes, we selected 6 sizes, deemed to be both likely to occur in real world problems and to exercise different allocator behaviour. The sizes we selected are 8, 64, 512, 4096, 16384 and 65536. For each pair of sizes $(x, y)$ there are four possible benchmarks to be run: $x$ allocated temporally first overflowing into $y$, $x$ allocated temporally first underflowing into $y$, $y$ allocated temporally first overflowing into $x$, and $y$ allocated temporally first underflowing into $x$. This produces 72 benchmarks to run for each combination of allocator (3), noise (3) and starting state (4), giving 2592 benchmarks in total.


\item For each source and destination combination size, we made available to the analyser an interaction sequence which triggers an allocation of the source size, an interaction sequence which triggers an allocation of the destination size, and interaction sequences for freeing each of the allocations.

\end{itemize}

\begin{figure}[t]
\centering
\includegraphics[scale=.45]{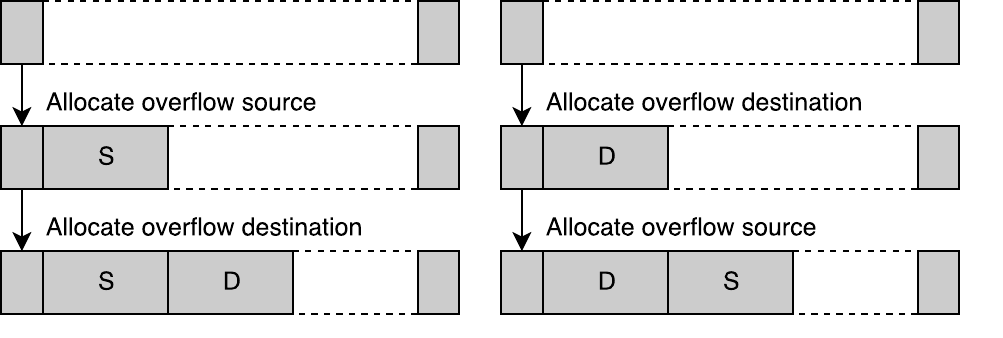}
\caption{For an allocator that splits chunks from the start of free blocks, the natural order, shown on the left, of allocating the source and then the destination produces the desired layout, while the reversed order, shown on the right, results in an incorrect layout.}
\label{fig:both_orders}
\end{figure}

\begin{figure}[t]
\centering
\includegraphics[scale=.45]{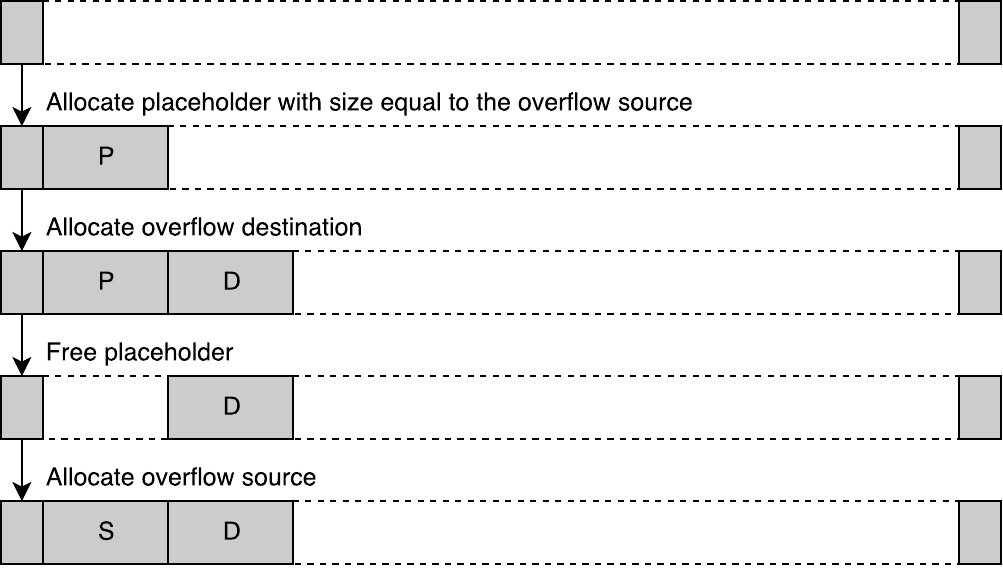}
\caption{A solution for the reversed allocation order to corruption direction relationship. A hole is created via a placeholder which can then be used for the source.}
\label{fig:reversed_solution}
\end{figure}

The $m$ and $r$ parameters to Algorithm~\ref{alg:search} were set to 1000 and .98 respectively\footnote{To determine reasonable values for these parameters, we constructed a small, distinct set of benchmarks explicitly for this purpose and separate to those used in our evaluation.}.The $g$ parameter was set to 500,000. A larger value would provide more opportunities for the search algorithm to find solutions, but with 2592 total benchmarks to run, and 500,000 executions taking in the range of 5-15 minutes depending on the number of interactions in the starting state, this was the maximum viable value given our computational resources. The results of the benchmarks averaged across all starting states can be found in Table~\ref{tab:synth_results}, with the full results in the appendices in Table~\ref{tab:full_synth}. 

To understand the `\% Natural' and `\% Reversed' columns in the results table we must define the concept of the \emph{allocation order to corruption direction relationship}. We refer to the case of the allocation of the source of an overflow temporally first, followed by its destination, or the allocation of the destination of an underflow temporally first, followed by its source as the \emph{natural} relationship. This is because most allocators split space from the start of free chunks and thus, for an overflow, if the source and destination are both split from the same chunk and the source is allocated first then it will naturally end up before the destination. The reverse holds for an underflow. We refer to the relationship as \emph{reversed} in the case of the allocation of the destination temporally first followed by the source for an overflow, or the allocation of the source temporally first followed by the destination for an underflow. We expect this case to be harder to solve for most allocators, as the solution is more complex than for the \emph{natural} relationship. A visualisation of this idea can be seen in Figure~\ref{fig:both_orders} and a solution for the reversed case is shown in Figure~\ref{fig:reversed_solution}.

From the benchmarks a number of points emerge:

\begin{itemize}
    \item When segregated storage is not in use, as with \dlmalloc and \avrlibc, and when there is no noise, 98\% to 100\% of the benchmarks are solved. 
    \item Segregated storage significantly increases problem difficulty. With no noise, the overall success rate drops to 72\% for \tcmalloc. 
    \item With the addition of a single noisy allocation, the overall success rate drops to close to 50\% across all allocators. 
    \item The order of allocation for the source and destination matters. A layout conforming to the natural allocation order to corruption direction relationship was easier to find in all problem instances. With four noisy allocations the success rate for problems involving the natural allocation order ranges from 44\% to 51\%, but drops to between 17\% and 38\% for the reversed order. It is also worth noting that the difference in success rate between natural and reversed problem instances is lower for \avrlibc than for \dlmalloc and \tcmalloc. This is because in some situations \avrlibc will split space from free chunks from the end instead of from the start. Thus, a reversed order problem can be turned into a natural order problem by forcing the heap into such a state, and this is often easier than solving the reversed order problem.
    \item We ran each experiment 9 times, and if all $9 * 500,000$ executions are taken together then 78\% of the benchmarks are solved \emph{at least} once. In other words, only 22\% of the benchmarks were never solved by our approach, which is quite encouraging given the simplicity of the algorithm.
\end{itemize}

\subsection{PHP-Based Benchmarks}
\label{sec:php_experiments}

To determine if automatic HLM is feasible in real world scenarios we selected three heap overflow vulnerabilities and ten dynamically allocated structures that were identified by \SHRIKE as being potentially useful targets (namely structures that have pointers as their first field). Pairing each vulnerability with each target structure provides a total of 30 benchmarks. For each, we ran an experiment in which \SHRIKE was used to search for an input which would place the overflow source and destination structure adjacent to each other. 

A successful outcome means the system can discover how to interact with the underlying allocator via \PHP's API, identify how to allocate sensitive data structures on the heap, and construct a \PHP program which places a selected data structure adjacent to the source of an OOB memory access. This saves an exploit developer a significant amount of effort, allowing them to focus on how to leverage the resulting OOB memory access.

Our evaluation utilised the following vulnerabilities:

\begin{itemize}
    \item \textbf{CVE-2015-8865}. An out-of-bounds write vulnerability in \texttt{libmagic} that exists in \PHP up to version 7.0.4.  
    \item \textbf{CVE-2016-5093}. An out-of-bounds read vulnerability in \PHP up to version 7.0.7, related to string processing and internationalisation.
    \item \textbf{CVE-2016-7126}. An out-of-bounds write vulnerability in \PHP up to version 7.0.10, related to image processing.
\end{itemize}

The ten target structures are described in the appendix in Table~\ref{tab:target_structures} and the full details of all 30 experiments can be found in Table~\ref{tab:php_cve_results}. As with the synthetic benchmarks, the $m$ and $r$ arguments to the \inlinecode{Search} function were set to 1000 and .98 respectively. Instead of limiting the number of executions via the $g$ parameter the maximum run time for each experiment was set to 12 hours. The following summarises the results:

\begin{itemize}
    \item \SHRIKE succeeds in producing a \PHP program achieving the required layout in 21 of the 30 experiments run and fails in 9 (a 70\% success rate). 
    \item There are 15 noise-free benchmarks of which \SHRIKE solves all 15, and 15 noisy benchmarks of which \SHRIKE solves 6. This follows what one would expect from the synthetic benchmarks. 
    \item In the successful cases the analysis took on average 571 seconds and 720,000 candidates. 
\end{itemize}

Of the nine benchmarks which \SHRIKE does not solve, eight involve CVE-2016-7126. The most likely reason for the difficulty of benchmarks involving this vulnerability is noise in the interaction sequences involved. The source buffer for this vulnerability results from an allocation request of size 1, which \PHP rounds up to 8 -- an allocation size that is quite common throughout \PHP, and prone to occurring as noise. There is a noisy allocation in the interaction sequence which allocates the source buffer itself, several of the interaction sequences which allocate the target structures also have noisy allocations, and all interaction sequences which \SHRIKE discovered for making allocations of size 8 involve at least one noisy allocation. For example, the shortest sequence discovered for making an allocation of size 8 is a call to \inlinecode{imagecreate(57, 1)} which triggers an allocation of size 7360, two allocations of size 8 and two allocations of size 57. In contrast, there is little or no noise involved in the benchmarks utilising CVE-2016-5093 and CVE-2015-8865.

\subsection{Generating a Control-Flow Hijacking Exploit for PHP}
\label{sec:generating_exploit}

To show that \SHRIKE can be integrated into the development of a full exploit we selected another vulnerability in \PHP. CVE-2013-2110 allows an attacker to write a NULL byte immediately after the end of a heap-allocated buffer. One must utilise that NULL byte write to corrupt a location that will enable more useful exploitation primitives. Our aim is to convert the NULL byte write into both an information leak to defeat ASLR and the ability to modify arbitrary memory locations.

We first searched \SHRIKE's database for interaction sequences that allocate structures that have a pointer as their first field. This lead us to the \inlinecode{imagecreate} function which creates a \inlinecode{gdImage} structure. This structure uses a pointer to an array of pointers to represent a grid of pixels in an image. By corrupting this pointer via the NULL byte write, and then allocating a buffer we control at the location it points to post-corruption, an attacker can control the locations that are read and written from when pixels are read and written. 

Listing~\ref{lst:exploit_template} shows the template provided to \SHRIKE. In less than 10 seconds \SHRIKE finds an input that places the source immediately prior to the destination. Thus the pointer that is the first field of the \inlinecode{gdImage} structure is corrupted. Listing~\ref{lst:php_sample_result} in the appendices shows part of the generated solution. After the corruption occurs the required memory read and write primitives can be achieved by allocating a controllable buffer into the location where the corrupted pointer now points. For brevity we leave out the remaining details of the exploit, but it can be found in full in the \SHRIKE repository. The end result is a \PHP script that hijacks the control flow of the interpreter and executes native code controlled by the attacker.

\subsection{Research Questions}

\textbf{RQ1: What factors most significantly impact the difficulty of the heap layout manipulation problem in a deterministic setting?}

The following factors had the most significant impact on problem difficulty:

\begin{itemize}
    \item \textbf{Noise.} In the synthetic benchmarks, noise clearly impacts difficulty. As more noise is added, more holes typically have to be created. In the worst case (\dlmalloc) we see a drop off from a 99\% overall success rate to 33\% when four noisy allocations are included. A similar success rate is seen for \avrlibc and \tcmalloc with four noisy allocations. In the evaluation on \PHP noise again played a significant role, with \SHRIKE solving 100\% of noise-free instances and 40\% of noisy instances. 

    \item \textbf{Segregated storage.} In the synthetic benchmarks segregated storage leads to a decline in the overall success rate on noise-free instances from 100-99\% to 72\%. 
    
    \item \textbf{Allocation order to corruption direction relationship.} For all configurations of allocator, noise and starting state, the problems involving the natural order were easier. For the noise-free instances on \avrlibc and \dlmalloc the difference is in terms of solved problems is just 1-2\%, but as noise is introduced the success rate between the natural and reversed benchmarks diverges. For \dlmalloc with four noisy allocations the success rate for the natural order is 49\% but only 17\% for the reversed order, a difference of 32\%.
\end{itemize}

\textbf{RQ2: Is pseudo-random search an effective approach to heap-layout manipulation?}

Without segregated storage, when there is no noise then 100-99\% of problems were solved, with most experiments taking 15 seconds or less. As noise is added the rate of success drops to 51\% and 46\% for a single noisy allocation, for \dlmalloc and \avrlibc respectively, and then to 41\% and 33\% for four noisy allocations. The extra constraints imposed on layout by segregated storage present more of a challenge. On noise-free runs the rate of success is 72\% and drops to 52\% and 37\% as one and four noisy allocations, respectively, are added. However, as noted in section~\ref{sec:synth_experiments}, if all 10 runs of each experiment are considered together then 78\% of the benchmarks are solved at least once.

On the synthetic benchmarks it is clear that the effectiveness of pseudo-random search varies depending on whether segregated storage is in use, the amount of noise, the allocation order to corruption direction relationship and the available computational resources. In the best case, pseudo-random search can solve benchmarks in seconds, while in the more difficult ones it still attains a high enough success rate to be worthwhile given its simplicity. 

When embedded in \SHRIKE, pseudo-random search approach also proved effective, with similar caveats relating to noise. 100\% of noise-free problems were solved, while 40\% of those involving noise were. On average the search took less than 10 minutes and 750,000 candidates, for instances on which it succeeded.

\textbf{RQ3: Can heap layout manipulation be automated effectively for real-world programs?}

Our experiments with \PHP indicate that automatic HLM can be performed effectively for real world programs. As mentioned in RQ2, \SHRIKE had a 70\% success rate overall, and a 100\% success rate in cases where there was no noise.

\SHRIKE demonstrates that it is possible to automate the process in an end-to-end manner, with automatic discovery of a mapping from the target program's API to interaction sequences, discovery of interesting corruption targets, and search for the required layout. Furthermore, \SHRIKE's template based approach show that a system with these capabilities can be naturally integrated into the exploit development process.

\subsection{Generalisability}

Regarding generalisability, our experiments are not exhaustive and care must be taken in extrapolating to benchmarks besides those presented. However, we believe that the presented search algorithm and architecture for \SHRIKE are likely to work similarly well with other language interpreters. \SHRIKE depends firstly on some means to discover language constructs and correlate them with their resulting allocator interactions, and secondly on a search algorithm that can piece together these fragments to discover a required layout. The approach used in \SHRIKE to solve the first problem is based on previous work on vulnerability detection that has been shown to work on interpreters for Javascript and Ruby, as well as PHP~\cite{holler12Langfuzz, heelan14ghosts}. Our extensions, namely a different approach to fuzzing as well as instrumentation to record allocator interactions, do not threaten the underlying assumptions of the prior work. Our solution to the second problem, namely the random search algorithm, has demonstrated its capabilities on a diverse set of benchmarks. Thus, we believe it is reasonable to expect similar results versus targets that rely on allocators with a similar architecture.

\subsection{Threats to Validity}

The results on our synthetic benchmarks are impacted by our choice of source and destination sizes. There may be combinations of these that produce layout problems that are significantly more or less difficult to solve. A different set of starting sequences, or available interaction sequences may also impact the results. We have attempted to mitigate these issues by selecting diverse  sizes and starting sequences, and allowing the analysis engine to utilise only a minimal set of interaction sequences. 

Our results on \PHP are affected by our choice of vulnerabilities and target data structures, and we could have inadvertently selected for cases that are outliers. We have attempted to mitigate this possibility by utilising ten different target structures and vulnerabilities in three completely different sub-components of \PHP. The restriction of our evaluation to a language interpreter also poses a threat if considering generalisability, as the available interaction sequences may differ in other classes of software. We have attempted to mitigate this threat by limiting the interaction sequences used to those that contain an allocation of a size equal to one of the allocation sizes found in the sequences which allocate the source and destination.  

\section{Related Work}
\label{sec:related_work}
Dullien~\cite{dullien17WeirdMachine} formalises the concept of an exploit as the process of setting up and programming a \emph{weird machine}. In the context of this formalisation, heap layout manipulation can be viewed as part of the process for producing the correct \emph{sane state} from which to transition to a \emph{weird state}.

The hacking and security communities have extensively published on reverse engineering heap implementations~\cite{Valasek12Windows8, Yason16Windows10}, leveraging weaknesses in those implementations for exploitation~\cite{Mandt11KernelPoolExploitation, McDonald09PracticalXPExploitation, Phantasmal05Maleficarium}, and heap layout manipulation for exploitation~\cite{Phrack:MaXX, Phrack:ArgpVLC}. There is also work on constructing libraries for debugging heap internals~\cite{Phrack:ArgpFF} and libraries which wrap an application's API to provide layout manipulation primitives~\cite{Sotirov07HeapFengShui}. Manually constructed solutions for heap layout manipulation in non-deterministic settings are also commonplace in the literature of the hacking and security communities~\cite{hay09flashexploit, blazakis10interp}.

Several papers~\cite{Avgerinos11AEG, Brumley08APEG, Heelan09Msc} have focused on the AEG problem. These implementations are based on symbolic execution and exclusively focus on exploitation of stack-based buffer overflows. More recently, as part of the DARPA Cyber Grand Challenge~\cite{DARPA16CGC} (CGC), a number automated systems~\cite{stephens16driller, forallsecure16mayhemblog, grammatech16cgc, trailofbits15cgc} were developed which combine symbolic execution and high performance fuzzing to identify, exploit and patch software vulnerabilities in an autonomous fashion. As with earlier systems, none of the CGC participants appear to specifically address the challenges of heap-based vulnerabilities. 

In~\cite{repel17} the authors present work on exploit generation for heap-based vulnerabilities that is orthogonal to ours. Using a driver program the system builds a database of conditions on the heap layout that, if met, would allow for corruption of heap metadata to be turned into a \emph{write-N} primitive~\cite{Phrack:MaXX}. To leverage these primitives in an exploit for a real program it is assumed that an input is provided for the program that results in the required heap layout prior to triggering the metadata corruption. In this paper we have demonstrated an approach to producing inputs that satisfy heap layout constraints, and thus could be used to process vulnerability triggers into inputs that meet the requirements of their system.

Vanegue~\cite{vanegue15heapsemantics} defines a calculus for a simple heap allocator and also provides a formal definition~\cite{vanegue13grandchallenge} of the related problem of automatically producing inputs which maximise the likelihood of reaching a particular program state given a non-deterministic heap allocator.   



\section{Conclusion}
\label{sec:conclusion}
In this paper we have outlined the heap layout manipulation problem as a distinct task within the context of automated exploit generation. We have presented a simple, but effective, algorithm for HLM in the case of a deterministic allocator and a known starting state, and shown that it can succeed in a significant number of synthetic benchmarks. We have also described an end-to-end system for HLM and shown that it is effective when used with real vulnerabilities in the \PHP interpreter. 

Finally, we have demonstrated how a system for automatic HLM can be integrated into exploit development. The directives provided by \SHRIKE allow the exploit developer to focus on the higher level concepts in the exploit, while letting \SHRIKE resolve heap layout constraints. To the best of our knowledge, this is a novel approach to adding automation to exploit generation, and shows how an exploit developer's domain knowledge and creativity can be combined with automated reasoning engines to produce exploits. Further research is necessary to expand on the concept, but we believe such human-machine hybrid approaches are likely to be an effective means of producing exploits for real systems. 


\section{Acknowledgements}
 We would like to thank the anonymous reviewers for their valuable feedback. We would also like to thank Thomas Dullien for his insightful discussions on this topic, and John Galea for his reviews and suggestions. This research was supported by ERC project 280053 (CPROVER) and the H2020 FET OPEN 712689 SC$^2$.

{\footnotesize \bibliographystyle{acm}
\bibliography{references}}

\clearpage

\appendix{}
\large\noindent\textbf{Appendix}\normalsize




\ctable[
    caption = {Summary of the heap initialisation sequences for synthetic benchmarks. All sequences were captured by hooking the \malloc, \free, \realloc and \calloc functions of the system allocator, except for \texttt{php-emalloc} which was captured by hooking the allocation functions of the custom allocator that comes with \PHP.},
    label = {tab:init_seq_summary}, pos = {ht}, botcap,
]{@{}llll@{}}{
}
{
    \FL
    Title       & \makecell{\# Allocator\\Interactions} & \makecell{\# Allocs} & \makecell{\# Frees}
    \ML
    php-emalloc     & 571   & 366  & 205  \NN
    php-malloc      & 15078 & 12714 & 2634  \NN
    python-malloc   & 6160  & 3710  & 2450 \NN
    ruby-malloc     & 70895 & 51827 & 19068 \NN
    \LL
}

\ctable[
    caption = {Target structures used in evaluating SHRIKE. Each has a pointer as its first field.},
    label = {tab:target_structures}, pos = {ht}, botcap,
]{@{}llll@{}}{
}
{
    \FL
    Type                & Size      & \makecell{Allocation\\Function}
    \ML
    gdImage             & 7360      & imagecreate \NN
    xmlwriter\_object    & 16       & xmlwriter\_open\_memory \NN
    php\_hash\_data       & 32      & hash\_init  \NN
    int *               & 8         & imagecreatetruecolor \NN
    Scanner             & 24        & date\_create \NN
    timelib\_tzinfo      & 160      & mktime \NN
    HashTable           & 264       & timezone\_identifier\_list \NN
    php\_interval\_obj    & 64        & unserialize \NN
    int *               & 40        & imagecreatetruecolor \NN
    php\_stream          & 232      & stream\_socket\_pair \NN
    \LL
}

\begin{lstlisting}[float=p, language=PHP, caption={Part of the solution discovered for using CVE-2013-2110 to corrupt the \inlinecode{gdImage} structure, which is the $1^{st}$ allocation made by \inlinecode{imagecreate} on line 11. Multiple calls are made to functions that have been discovered to trigger the desired allocator interactions. Frees are triggered by destroying previously created objects, as can be seen with \texttt{var\_shrike\_3} on line 14. The overflow source is the $1^{st}$ allocation performed by \inlinecode{quoted\_printable\_encode} on line 17}, label={lst:php_sample_result}]
<?php
$quote_str = str_repeat("\xf4", 123);

$var_vtx_0 = str_repeat("747 X ", 58);
$var_vtx_1 = str_repeat("747 X ", 58);
$var_vtx_2 = str_repeat("747 X ", 58);
$var_vtx_3 = imagecreatetruecolor(346, 48);
<...>
shrike_record_alloc(0, 1);
$image = imagecreate(1, 2);
<...>
$var_vtx_300 = str_repeat("747 X ", 58);
$var_vtx_3 = 0;
<...>
shrike_record_alloc(0, 2);
quoted_printable_encode($quote_str);
$distance = shrike_get_distance(1, 2);
if ($distance != 384) {
	exit("Invalid layout.\n");
}
\end{lstlisting}


\ctable[
    caption = {Synthetic benchmark results. For each experiment the search was run for a maximum of 500,000 candidates. All experiments were run 9 times and the results below are the average of those runs. `\% Solved' is the percentage of the 72 experiments for each row in which an input was found placing the source and destination adjacent to each other. `\% Natural' is the percentage of the 36 natural allocation order to corruption direction experiments which were solved. `\% Reversed' is the percentage of the 36 reversed allocation order to corruption direction experiments which were solved.},
    label = {tab:full_synth},
    star
]{@{}lllllllll@{}}{
}
{
    \FL
    Allocator       & Start State   & Noise & \makecell{\% Solved}  & \makecell{\% Natural} & \makecell{\% Reversed} 
    \ML
    avrlibc-r2537 & php-emalloc & 0 & 100 & 100 & 100 \NN
	avrlibc-r2537 & php-malloc & 0 & 100 & 100 & 100 \NN
	avrlibc-r2537 & python-malloc & 0 & 100 & 100 & 100 \NN
	avrlibc-r2537 & ruby-malloc & 0 & 99 & 100 & 98 \NN
	dlmalloc-2.8.6 & php-emalloc & 0 & 99 & 100 & 99 \NN
	dlmalloc-2.8.6 & php-malloc & 0 & 100 & 100 & 100 \NN
	dlmalloc-2.8.6 & python-malloc & 0 & 99 & 100 & 97 \NN
	dlmalloc-2.8.6 & ruby-malloc & 0 & 99 & 100 & 98 \NN
	tcmalloc-2.6.1 & php-emalloc & 0 & 73 & 79 & 67 \NN
	tcmalloc-2.6.1 & php-malloc & 0 & 77 & 80 & 75 \NN
	tcmalloc-2.6.1 & python-malloc & 0 & 63 & 63 & 62 \NN
	tcmalloc-2.6.1 & ruby-malloc & 0 & 75 & 78 & 71 \NN
	avrlibc-r2537 & php-emalloc & 1 & 55 & 51 & 59 \NN
	avrlibc-r2537 & php-malloc & 1 & 51 & 46 & 56 \NN
	avrlibc-r2537 & python-malloc & 1 & 49 & 51 & 46 \NN
	avrlibc-r2537 & ruby-malloc & 1 & 49 & 50 & 48 \NN
	dlmalloc-2.8.6 & php-emalloc & 1 & 49 & 65 & 32 \NN
	dlmalloc-2.8.6 & php-malloc & 1 & 49 & 62 & 37 \NN
	dlmalloc-2.8.6 & python-malloc & 1 & 42 & 56 & 27 \NN
	dlmalloc-2.8.6 & ruby-malloc & 1 & 43 & 58 & 27 \NN
	tcmalloc-2.6.1 & php-emalloc & 1 & 52 & 59 & 45 \NN
	tcmalloc-2.6.1 & php-malloc & 1 & 55 & 61 & 48 \NN
	tcmalloc-2.6.1 & python-malloc & 1 & 50 & 52 & 48 \NN
	tcmalloc-2.6.1 & ruby-malloc & 1 & 53 & 61 & 44 \NN
	avrlibc-r2537 & php-emalloc & 4 & 43 & 44 & 42 \NN
	avrlibc-r2537 & php-malloc & 4 & 40 & 41 & 40 \NN
	avrlibc-r2537 & python-malloc & 4 & 42 & 47 & 37 \NN
	avrlibc-r2537 & ruby-malloc & 4 & 39 & 45 & 33 \NN
	dlmalloc-2.8.6 & php-emalloc & 4 & 34 & 51 & 16 \NN
	dlmalloc-2.8.6 & php-malloc & 4 & 31 & 44 & 17 \NN
	dlmalloc-2.8.6 & python-malloc & 4 & 33 & 50 & 16 \NN
	dlmalloc-2.8.6 & ruby-malloc & 4 & 35 & 51 & 20 \NN
	tcmalloc-2.6.1 & php-emalloc & 4 & 40 & 53 & 27 \NN
	tcmalloc-2.6.1 & php-malloc & 4 & 39 & 53 & 25 \NN
	tcmalloc-2.6.1 & python-malloc & 4 & 32 & 42 & 22 \NN
	tcmalloc-2.6.1 & ruby-malloc & 4 & 38 & 54 & 22 \NN
    \LL
}

\ctable[
    caption = {Results of heap layout manipulation for vulnerabilities in \PHP. Experiments were run for a maximum of 12 hours. All experiments were run 3 times and the results below are the average of these runs. `Src. Size' is the size in bytes of the source allocation. `Dst. Size' is the size in bytes of the destination allocation. `Src./Dst. Noise' is the number of noisy allocations triggered by the allocation of the source and destination. `Manip. Seq. Noise' is the amount of noise in the sequences available to SHRIKE for allocating and freeing buffers with size equal to the source and destination. `Initial Dist.' is the distance from the source to the destination if they are allocated without any attempt at heap layout manipulation. `Final Dist.' is the distance from the source to the destination in the best result that SHRIKE could find. A distance of 0 means the problem was solved and the source and destination were immediately adjacent. `Time to best` is the number of seconds required to find the best result. `Candidates to best` is the number of candidates required to find the best result.},
    label = {tab:php_cve_results},
    notespar,
    star,
]{@{}llllllllllll@{}}{
}   
{
    \FL
    CVE ID       & \makecell{Src.\\Size} & \makecell{Dst.\\Size} & \makecell{Src./Dst.\\Noise} & \makecell{Manip. Seq.\\Noise} & \makecell{Initial\\ Dist.} & \makecell{Final\\Dist.} & \makecell{Time to\\Best} & \makecell{Candidates to\\Best}
    \ML
    2015-8865   & 480   & 7360  & 0     & 0     & -16384    & 0     & $<$1     & 106       \NN 
    2015-8865   & 480   & 16    & 0     & 0     & -491424   & 0     & 170     & 218809     \NN 
    2015-8865   & 480   & 32    & 0     & 0     & -96832    & 0     & 217    & 286313         \NN 
    2015-8865   & 480   & 8     & 0     & 1     & -540664   & 0     & 642    & 862689         \NN 
    2015-8865   & 480   & 24    & 0     & 0     & -151456   & 0     & 16     & 13263         \NN 
    2015-8865   & 480   & 160   & 0     & 0     & -57344    & 0     & $<$1     & 63         \NN 
    2015-8865   & 480   & 264   & 0     & 0     & -137344   & 0     & $<$1     & 84         \NN 
    2015-8865   & 480   & 64    & 1     & 0     & -499520   & 0     & 12    & 13967         \NN 
    2015-8865   & 480   & 40    & 0     & 0     & -128832   & 0     & 25     & 15113         \NN 
    2015-8865   & 480   & 232   & 0     & 0     & -101376   & 0     & $<$1     & 69         \NN 
    
    2016-5093   & 544   & 7360  & 1     & 0     & 84736     & 0     & $<1$      & 640          \NN 
    2016-5093   & 544   & 16    & 0     & 0     & -402592   & 0     & 4202      & 5295968 \NN
    2016-5093   & 544   & 32    & 0     & 0     & -7776     & 0     & 2392      & 3014661     \NN 
    2016-5093   & 544   & 8     & 0     & 1     & -406776   & 8     & 6905      & 9049924     \NN 
    2016-5093   & 544   & 24    & 0     & 0     & -62624    & 0     & 202       & 231884     \NN 
    2016-5093   & 544   & 160   & 0     & 0     & 80640     & 0     & $<1$      & 104     \NN 
    2016-5093   & 544   & 264   & 0     & 0     & -27712    & 0     & $<1$      & 76     \NN 
    2016-5093   & 544   & 64    & 1     & 0     & -410624   & 0     & 487       & 607824         \NN 
    2016-5093   & 544   & 40    & 0     & 0     & -31648    & 0     & 15        & 458         \NN 
    2016-5093   & 544   & 232   & 0     & 0     & 77312     & 0     & 3         & 116         \NN 
    
    2016-7126   & 1     & 7360  & 4     & 2     & 495576    & 0     & 958       & 1181098 \NN 
    2016-7126   & 1     & 16    & 0     & 4     & 4360      & 88    & 4816      & 6260800     \NN
    2016-7126   & 1     & 32    & 1     & 1     & 398808    & 64    & 5594      & 7272200     \NN 
    2016-7126   & 1     & 8     & 3     & 2     & -32       & 0     & 2662      & 3356935     \NN 
    2016-7126   & 1     & 24    & 3     & 1     & 344152    & 56    & 4199      & 5458700     \NN 
    2016-7126   & 1     & 160   & 14    & 1     & 483288    & 24    & 3005      & 3864430     \NN 
    2016-7126   & 1     & 264   & 0     & 1     & 379064    & 24    & 5917      & 7615179     \NN 
    2016-7126   & 1     & 64    & 1     & 3     & -3912     & 72    & 2752      & 3539072         \NN 
    2016-7126   & 1     & 40    & 5     & 1     & 375248    & 144   & 7980      & 10134600         \NN 
    2016-7126   & 1     & 232   & 0     & 1     & 439288    & 40    & 5673      & 7908162         \NN 
    \LL
}

\end{document}